\documentclass[11pt]{article}

\usepackage{acl}

\usepackage{times}
\usepackage{latexsym}
\usepackage[T1]{fontenc}
\usepackage[utf8]{inputenc}
\usepackage{microtype}
\usepackage{inconsolata}
\usepackage{graphicx}
\usepackage{booktabs}
\usepackage{array}
\usepackage{ragged2e}
\usepackage{enumitem}
\usepackage{amsmath}
\usepackage{amsfonts}
\usepackage{amssymb}
\usepackage{tikz}
\usetikzlibrary{arrows.meta, positioning, shapes.geometric, fit}

\title{Position: Scaling Human-AI Coding Collaboration Requires a Governable Consensus Layer}

\author{
Tianfu Wang$^{1}$ \quad Zhezheng Hao$^{2}$ \quad Yin Wu$^{1}$ \quad Wei Wu$^{1}$ \quad Haowen Yang$^{1}$ \\
{\bfseries Linxiao Gong$^{1}$ \quad Qiang Lin$^{3}$ \quad Hande Dong$^{3,\ast}$ \quad Nicholas Jing Yuan$^{1}$ \quad Hui Xiong$^{1,\ast}$} \\
\normalfont
$^{1}$The Hong Kong University of Science and Technology (Guangzhou) \\
$^{2}$Zhejiang University \quad
$^{3}$Tencent \\
\small \texttt{tianfuwang.cs@gmail.com}\quad $^{\ast}$Corresponding authors: \texttt{donghd66@gmail.com}; \texttt{xionghui@ust.hk}
}

\hypersetup{
  pdftitle={Position: Scaling Human-AI Coding Collaboration Requires a Governable Consensus Layer},
  pdfauthor={Tianfu Wang, Zhezheng Hao, Yin Wu, Wei Wu, Haowen Yang, Linxiao Gong, Qiang Lin, Hande Dong, Nicholas Jing Yuan, Hui Xiong},
  pdfsubject={A position paper on governable project-state consensus for human and AI coding collaboration},
  pdfkeywords={human and AI collaboration, coding agents, project-state review, agentic consensus}
}

\begin{document}
\maketitle

\begin{abstract}
Vibe coding turns natural-language intent into working software, but code generation alone does not give humans and agents a shared account of what they are building.
As collaboration spans participants, components, and releases, design choices must remain understandable and open to revision.
We argue that scaling human and AI coding collaboration requires a governable consensus layer.
\textbf{Agentic Consensus} organizes collaboration around a versioned project-state model linking selected intent, structural commitments, executable artifacts, and evidence.
Participants can clarify intent before implementation and coordinate proposed changes against commitments that humans can inspect and revise.
Bidirectional synchronization connects intended changes to implementation and exposes conflicts or stale evidence as artifacts evolve.
We implemented its review workflow in Consensus Control.
In a small exploratory human study, 10 participants used the system on projects from their everyday work and reported clearer, faster understanding of agent-submitted changes.
A three-condition comparison with the same participants collected 60 responses, with near-ceiling decision scores and measured representation and pipeline costs.
Our research agenda asks how shared project state can preserve human understanding and control as agent contributions accumulate.
\end{abstract}

\section{Introduction}
Vibe coding~\cite{karpathy2025vibecoding} lets users develop software by expressing intent, inspecting results, and asking agents for revisions.
In a continuing project, the challenge extends beyond producing the next working version: people and agents must understand the system's structure and coordinate changes as requirements evolve.
Design choices need to remain available across contributors, coding sessions, and releases.

Consider a team that asks coding agents to ``make long-running tools asynchronous without breaking existing assistants.''
One agent returns a job identifier, another makes an SDK method syntactically asynchronous while preserving blocking behavior, and a third makes the planner expect callbacks.
Each patch passes its local tests, yet the deployed versions no longer share one protocol.
The request admitted several interpretations; no shared record identified the selected meaning, supported consumers, or evidence needed for cutover.
The missing decision is whether completion means returning a result, accepting a job, or delivering an event, and how old clients should behave during the transition.
Prompt underspecification can leave such choices to fragile defaults~\cite{yang-etal-2026-prompts}.

\begin{figure}[t]
  \centering
  \includegraphics[width=\columnwidth]{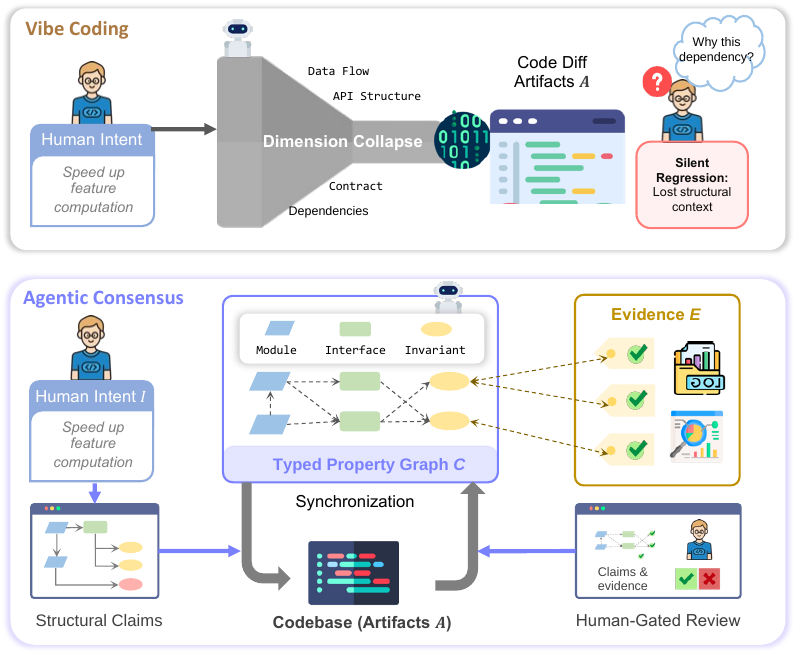}
  \caption{Vibe coding (top) maps natural-language intent directly to code, risking loss of the structural context needed for review. Agentic Consensus (bottom) makes those commitments explicit in a versioned project-state model $C$, illustrated here as a typed property graph, linked bidirectionally to executable artifacts $A$ and claim-level evidence $E$ for human-gated review.}
  \label{fig:comparison}
\end{figure}

We call this the \emph{representation gap}: executable changes do not, by themselves, provide a shared account of the project's commitments.
Code records what was implemented, while rationale remains scattered across prompts, tickets, documentation, and collaborators' memory.
RECAP links prompts and code edits to reconstruct programming interactions~\cite{he2026recap}; project-state review additionally asks which commitments remain authorized across releases.
A reviewer must reconstruct what was intended, which dependencies matter, and whether earlier evidence still applies.
A diff can reveal the new return value without establishing whether older callers may rely on the previous one. That obligation belongs to the evolving project, not to any single patch.

Our position is that \emph{scaling human and AI coding collaboration requires a governable consensus layer}: a shared project state that participants can inspect, negotiate, and revise.
For structurally consequential changes, this requires project-state review alongside code review.
The governed path applies when a change affects an independently consumed interface; ownership, authority, or policy; a migration or compatibility window; or an evidence obligation that depends on state outside the diff.
Ambiguous cases require clarification; other local edits stay on the ordinary diff-and-test path.

\textbf{Agentic Consensus} organizes human and agent collaboration around a versioned project-state model $C$ (Figure~\ref{fig:comparison}).
It binds a request and its selected interpretation to stable claims, artifact revisions, applicable evidence, unresolved conflicts, and a human decision.
The consensus state records the commitments that authorized participants have accepted and the conditions under which they remain applicable.
Before implementation, participants can compare interpretations and clarify how a request changes those commitments.
Agents propose changes against this shared state; humans inspect how those changes affect commitments and what evidence supports them.
When artifacts change, the same claims must be revisited: supporting evidence may become stale, competing proposals may conflict, and a previously acceptable exception may no longer apply.

Participants should be able to follow one commitment from the original request to the current implementation, see which evidence supports it, and identify who can authorize a revision.
The same record supports clarifying a request, coordinating an implementation, reviewing a patch, and revisiting an earlier choice.

\paragraph{Contributions.} We make three contributions:
\begin{itemize}[leftmargin=*, itemsep=2pt]
  \item A project-state view of human and AI collaboration, with a routing rule for structurally consequential changes requiring project-state review.
  \item A governance contract linking intent, artifacts, evidence, and authority, instantiated in the Consensus Control review workspace.
  \item An information-matched comparison of three review representations with 10 coding-agent users, combining field use with decision-quality, workload, and cost measurements.
\end{itemize}

\section{Related Work}

\textbf{Structural Representations for Code.}
Code representations include narratives, graphs, and bidirectional mappings \cite{knuth1984,luo2025xkg,czarnecki2009bidirectional,yamaguchi2014cpg,chen2025cgbridge,chivukula2025agint}; Knowledge Objects retain claims and validation metadata \cite{liu2026knowledgeobjects}. Requirements traceability follows requirements from their origins to implementation~\cite{gotel1994traceability}; we bind those links to versioned evidence and decisions.

\textbf{Trust in Agentic Software Engineering.}
Formal methods, SAFE AI, and coding agent studies motivate auditable state and expose acceptance and refactoring gaps \cite{clarke1999,navneet2025safeai,li2025aidev,horikawa2025refactoring}. Related work examines coordination in agent teams~\cite{hong2024metagpt,qian2024chatdev,hao2026evolveteam}, concurrency control and contracts~\cite{yang2026concurrency,kim2026contracts}, and ambiguity in code review~\cite{lin2025codereviewqa}. RTADev aligns agents through checks and group review~\cite{liu2025rtadev}; our focus is human-authorized commitments across releases.

\textbf{Human and AI Coding Benchmarks.}
SWE-bench evaluates single-issue repository fixes \cite{jimenez2024swebench}, while FeatureBench exposes the difficulty of multi-commit feature development \cite{zhou2026featurebench}.
CentaurEval measures the value of human intervention on collaboration-necessary coding tasks \cite{luo2026centaureval}.
Evaluating our position additionally requires longitudinal project state and information-matched review conditions that distinguish violation identification, evidence adequacy, policy-compatible disposition, and authorization compliance.

\section{The Agentic Consensus Contract}

Agentic Consensus makes project commitments explicit in a versioned model $C$: the shared state used to clarify intent, coordinate proposals, and support human-authorized revisions.
The two illustrative traces motivate the contract's components. In the protocol migration, language provenance records which meaning of ``asynchronous'' was selected. Stable claims and artifact anchors identify which release implements that choice. In the authority-drift scenario, evidence remains tied to artifact versions, so an old passing trace cannot validate a new connector. Conflict and authority records preserve the scope of exceptions when proposals overlap. In both cases, a joint proposal and decision package binds approval to the exact state transition under review. Realization and rehydration may be manual or automated, but must record correspondence and unresolved gaps in both directions.

\subsection{Consensus Model}

\subsubsection{Versioned Project State}
We denote the committed workflow state at version $t$ by $S^t=(I^t,C^t,A^t,E^t)$:
\begin{itemize}[leftmargin=*, itemsep=2pt]
  \item $I^t$: user intent, potentially partial, implicit, and time-varying,
  \item $C^t$: the versioned project-state model (schema, graph, and review views),
  \item $A^t$: realized artifacts that execute in the world (code, configs, and dataflows), and
  \item $E^t$: evidence linked to claims in $C^t$ (tests, traces, and provenance).
\end{itemize}
The superscript identifies a committed version; uncommitted proposals are represented as deltas against that version.
Here, \emph{consensus} means the currently authorized, versioned project state rather than agent voting, unanimity, or epistemic truth. It may still contain contested claims or scoped waivers, provided that disagreement and exception state remain explicit.
Reviewers use $C$ to inspect intent, artifacts, and evidence together and to govern changes to them.
$I$ enters the system through propose/clarify operations that encode intent into structural claims within $C$.
$A$ connects to $C$ via the realization operator $\Phi$, which gives distributed project-level rationale stable claims and auditable artifact anchors.
$E$ attaches to claims in $C$ as validation signals; stable claim identity and cross-version provenance make test results interpretable as persistent evidence for or against project-level commitments.
Provenance models link entities, activities, and responsible agents~\cite{moreau2013prov}.
$C$ organizes intent, code, and tests into a shared review surface while each retains its original role.

\subsubsection{Intent and Clarification}
We model language-mediated intent as a set of candidate state deltas:
\begin{equation}
\Omega(I^t,C^t)=\{(\Delta C_j,\eta_j,\rho_j)\}_{j=1}^{m},
\end{equation}
where each $\Delta C_j$ is a candidate structural interpretation, $\eta_j$ records unresolved uncertainty, and $\rho_j$ links request spans and interaction context to candidate claims. A clarification operation $\kappa(\Omega,h)\rightarrow\Omega'$ incorporates a human answer or policy constraint $h$ to eliminate, refine, or add candidates. For ``asynchronous without breaking existing assistants,'' the disputed slots are completion semantics, consumer scope, and compatibility duration. A discriminating question asks whether a call returns a result or a durable job identifier, and which released consumers require an adapter. Its answer changes the candidate contract before code generation. The selected interpretation retains the original request, material alternatives, clarification history, and selecting actor. Humans or agents may supply candidates; candidate completeness and confidence calibration remain separate evaluation targets. Selection authority belongs to the designated human or policy process.

\subsubsection{Auditable Graph}
The consensus layer is a typed, queryable schema with multiple synchronized projections.
Our reference design uses a labeled property graph whose nodes represent modules, interfaces, data entities, and invariants, and whose edges represent dependencies, contracts, and temporal relations.
The schema is representation-agnostic: a ledger or integrated Git/CI workflow can implement the same claim identities, artifact links, evidence lifecycle, and authorization rules. We use a graph to expose dependencies and their paths during review.

Evidence records in $C$ link directly to the claims they support or challenge.
For example, an edge ``dataset $D$ conforms to schema $S$'' should have pointers to validation jobs and their timestamps.
This auditability requires a precise correspondence between graph elements and artifacts:
\begin{itemize}[leftmargin=*, itemsep=2pt]
  \item \textit{Structural mapping:} nodes map to code units (packages, classes, functions), configuration blocks, or dataset definitions.
  \item \textit{Behavioral mapping:} edges map to explicit call sites in code, API contracts, data-lineage relations, or testable invariants.
  \item \textit{Round-trip integrity:} changes in $C$ compile down to edits in $A$; changes in $A$ rehydrate back into $C$ through static analysis, runtime traces, and provenance logs tied to artifact revisions.
\end{itemize}

\subsubsection{Structural Sandbox}
$C$ is where agents reason about structure: manipulating dependencies, running analyses, and evaluating counterfactuals without modifying artifacts.

For instance, the system should answer questions of the form:
\begin{quote}
If I delete this node in the consensus graph, which dependencies will break in the downstream code?
\end{quote}

Operationally, this requires that $C$ encodes enough dependency information (and links to evidence) to support ``what-if'' queries and rollback-safe planning.
If the answer depends on unknown or implicit dependencies, the system should expose competing candidate changes and request discriminating evidence before selection.
Task-specific subgraphs expose the dependencies relevant to the current review, so reviewers can inspect affected paths without navigating the entire project graph.

\subsection{Round-Trip Synchronization}
As commitments and artifacts change, synchronization must preserve their correspondence and update the evidence links used in review.

\subsubsection{Realization and Rehydration}
A session alternates between consensus moves (proposing or modifying $C$) and realization moves (editing $A$ and rehydrating back into $C$). Synchronization operators couple these moves, preserving round-trip integrity and updating evidence links.
Let $\widehat A$ denote the candidate artifact target, identified by an immutable revision independently of governed version $t$. We define two synchronization operators:
\begin{itemize}[leftmargin=*, itemsep=2pt]
  \item $\Phi(C^t,A^t,\Delta C) \rightarrow (\widehat{\Delta A},O_{\Phi})$ (\emph{realize}): produce a candidate artifact diff and expose claims or preconditions that could not be realized.
  \item $\Psi(C^t,A^t,\widehat A) \rightarrow \{(\widehat{\Delta C}_k,U_k)\}_{k=1}^{K}$ (\emph{rehydrate}): produce candidate project-state deltas and record relations that remain unobservable or ambiguous.
\end{itemize}
In practice, $\Psi$ can combine static analysis (e.g., AST) \cite{yamaguchi2014cpg}, dynamic invariant detection \cite{ernst2007daikon}, data lineage/provenance, and test results. Relations that these methods cannot observe must remain marked as uncertain in the recovered project state.

\subsubsection{Monitoring Consistency}
Round-trip consistency is a monitored approximation: $\Phi$ may discard structural information with no artifact-level counterpart, and $\Psi$ is underdetermined.
Let $\widehat{\Delta C}_{k^\star}$ be the candidate selected after rehydrating the realized artifact change. A viable implementation should monitor the soft invariant $D_{\mathrm{rt}}=d_C(C^t\!\oplus\!\Delta C,\,C^t\!\oplus\!\widehat{\Delta C}_{k^\star})$ with discrepancy and observable coverage as explicit review signals.
This permits incremental adoption: partial layers can be evaluated before full round-trip fidelity.
Low round-trip discrepancy is an internal consistency signal; independent invariants and policy checks evaluate semantic correctness because shared errors in $\Phi$ and $\Psi$ can preserve the same claim.
Appendix~\ref{app:state-contract} specifies the extended operator outputs and versioned change package.

For ambiguous deltas, implementations run discriminating checks before requesting human clarification. Escalation thresholds require calibration against interruption cost and the risk of attributing the wrong intent to a change.

\subsubsection{Evidence-Gated Change Control}
Every proposal package names its base version, selected interpretation and language provenance, affected subgraph, predicted artifact change $\Delta A$, evidence obligations, unresolved relations, conflicts, and proposed waivers.
The reviewer inspects those structural claims together with the code diff and available evidence: $C$ organizes the review, while executable artifacts and observed evidence provide the underlying operational record.
If the committed state has advanced or overlapping proposals modify the same claims, the system must expose the conflict and require rebase or comparison before update.
A separate decision record binds the proposal to a disposition, governing policy, authorized actor, rationale, final waivers, and atomic provenance. An admitted proposal plus this decision record forms a committed package; rejected or rebased proposals remain auditable and leave $C$ unchanged. Rollback restores prior correspondence, retains historical evidence, and recomputes its applicability to the restored state. Appendix~\ref{app:state-contract} formalizes these objects.

Consensus Control executes this state machine with persistent versioned records, actor-bound project authority, admission checks, and rollback. It also runs one bounded $\Phi$/$\Psi$ path for Python import substitution, from an isolated proposal commit through independent diff rehydration and discrepancy reporting; Appendix~\ref{app:prototype} and Table~\ref{tab:runtime-coverage} report the implemented coverage.

The reference system records the local proposal, decision, audit, and project-state update in one atomic transition. It does not merge pull requests or deploy code. Production integrations must bind admission to an immutable target, status checks, merge receipt, and post-merge rehydration; these remain distinct, provenance-linked events.

\subsection{Role-Agnostic Governance}
Maintaining $C$ spans intent elicitation, artifact editing, evidence generation, conflict detection, and task-specific presentation. One agent may perform these functions, or a pipeline may decompose them as in multi-agent software-development systems \cite{hong2024metagpt,qian2024chatdev}. Across agent configurations, every proposal uses the same versioned change package before a human decides to commit, waive, rebase, or roll back. Appendix~\ref{app:roles} shows how these functions can be assigned to distinct agent roles.

\section{Illustrative Governance Traces}
\begin{figure*}[t]
  \centering
  \includegraphics[width=0.96\textwidth]{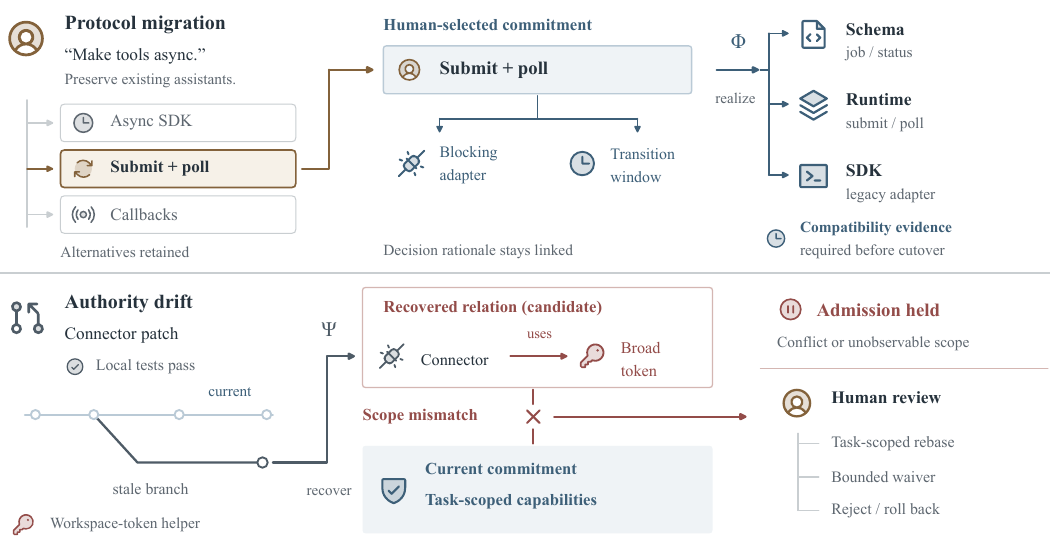}
  \caption{Two synthetic governance traces. \textbf{Top:} A human selects among competing interpretations before $\Phi$ proposes linked artifact changes; cutover requires compatibility evidence. \textbf{Bottom:} $\Psi$ recovers a candidate broad-credential relation from a stale-branch patch. Comparison with current task-scoped authority holds admission for human disposition; unobservable scope also prevents admission by default.}
  \label{fig:democases}
  \vspace{-12pt}
\end{figure*}

We use two hypothetical, multi-release scenarios in which language-mediated coding is part of the failure mechanism (Figure~\ref{fig:democases}).
The first traces ambiguous intent into incompatible implementations; the second traces a locally valid patch back to a violated authority commitment.
Each trace identifies what a reviewer needs to know beyond the local diff to decide whether the change can proceed.

\subsection{Ambiguous Protocol Migration}

\textit{Project setting.} A tool-using agent platform supports long-running operations across schemas, planner prompts, runtimes, SDKs, connectors, and evaluation harnesses.
A human asks coding agents to ``make long-running tools asynchronous without breaking existing assistants.''
The under-specified term \emph{asynchronous} may mean a nonblocking SDK call, a persistent submission and polling protocol, or callback events, each with different retry, compatibility, and rollback semantics.

\textit{Divergent interpretations.} Agents working in different repositories or from different base versions select different interpretations.
One returns a job identifier, another preserves a blocking protocol behind an \texttt{async} SDK method, and a third teaches the planner to await callbacks.
Each patch can satisfy its local tests while mixed deployed versions disagree about when a call has completed and how its result is recovered.
Product authorization still requires an accountable decision selecting the protocol and its supported consumers.

\textit{Prospective control in $C$.} Before realization, $C$ represents the alternatives as competing $\Delta C$s with affected consumers.
A human commits a submit-and-poll protocol with a temporary blocking adapter, together with the compatibility window, idempotency rule, deprecation condition, and rollback point.
$\Phi$ proposes linked artifact changes; protocol conformance, mixed-version replay, and consumer coverage remain attached to the claims they support.
If a later patch preserves an undeclared blocking caller, $\Psi$ surfaces a candidate conflict and blocks an implicit redefinition of ``asynchronous.''
A later request to ``restore callbacks'' must distinguish callbacks rejected as completion semantics from optional notifications omitted from an implementation. The current polling specification alone leaves that history open; the selected alternative and clarification rationale answer it. A versioned specification already preserving that lineage meets the same obligation.

\subsection{Authority Drift}

\textit{Project setting.} A platform migrates tool execution from workspace-wide credentials to task-scoped capabilities.
The commitment spans tool descriptions, planner policy, runtime validation, a secrets broker, connector SDKs, and audit traces.
Legacy credentials remain available only to named connectors during a bounded transition; the intended invariant is that an agent can exercise no capability beyond those granted for its current task.

\textit{Unintended authority expansion.} Several releases later, a coding agent working from a stale branch adds a connector by reusing a convenient workspace-token helper.
Repository tests pass because their mock authorizes every tool, and the connector works for the requested task.
Yet the patch expands the deployed agent's action space: policy compliance additionally requires the prompts, schemas, and runtime enforcement to respect the same task-scoped authority.

\textit{Retrospective control in $C$.} $C$ versions the authority invariant, its artifact anchors, the transitional exception, and the evidence required to retire it.
$\Psi$ proposes a \texttt{uses-broad-credential} relation and compares it with the committed task scope; unresolved dynamic credential resolution is recorded as \emph{unobservable} and blocks safe admission by default.
The reviewer sees the implicated artifacts, exception expiry, and current evidence together, then requires a task-scoped rebase, grants a time-bounded waiver, rejects the patch, or rolls it back.
Policy checks and denied-capability traces remain bound to the artifact versions they validate.

\paragraph{Design implications.}
The first trace exercises prospective synchronization, $I\rightarrow\Delta C\rightarrow\Phi\rightarrow\Delta A$: a human selects among semantically plausible interpretations before agents propagate one into artifacts.
The second exercises retrospective synchronization, $\Delta A\rightarrow\Psi\rightarrow\widehat{\Delta C}$: a locally valid patch is compared with a persistent, human-authorized commitment.
Product semantics and policy remain under authorized human decision; agent agreement and graph extraction supply proposals and evidence.
In both traces, the reviewer needs to recover a decision made in an earlier release and check whether the proposed artifacts still respect it.

\section{Consensus Control}
\label{sec:system}

We implemented Consensus Control as an authenticated local, multi-user review workspace.\footnote{Anonymous repository: \url{https://anonymous.4open.science/r/agentic-consenus/}.} It registers a Git repository and immutable base and target revisions, and constructs a versioned review package with claim deltas, artifact paths, evidence records, unresolved findings, waivers, and an append-only decision log. A versioned Intent Map retains the request, candidate interpretations, clarification revisions, and explicit human selection. The map records uncertainty and request-to-claim provenance alongside each candidate. Clarification history and selection identify how the proposed commitment follows from the original request.

\subsection{Review Workspace}
The project workspace organizes review around commitments. Selecting a commitment brings its proposed meaning, source-linked readings, and implementing code into one view. The source inspector opens the version-pinned artifact alongside the claim, so a reviewer can check the interpretation against the code it names. Authored readings state how the code appears to support the claim and what still needs inspection; they do not establish the behavior of the running system.

The change map provides a complementary overview of the patch. Directory groups and file tiles expose the changed artifacts, while links connect the map to the review's claims and evidence. Tile area represents changed lines, giving a size overview rather than a risk score.

The evidence workspace organizes outstanding requirements by commitment. Each requirement exposes what must be checked and the actions for attaching evidence or recording a waiver. Recorded results, missing checks, and review findings remain distinguishable, so the presence of an evidence record does not by itself satisfy an obligation.

\subsection{Recorded Walkthrough}
Figure~\ref{fig:reviewer-prototype} presents a persisted review of a question-game application with 17 changed files, three linked claims, and three evidence records. Its first proposed commitment requires exactly three hints per question, revealed in order. The workspace places three authored source readings beside this statement: the generator fills missing hints but returns extras unchanged; the page renders the returned array and uses the hint count to control unlocking; and the hint component defines three labels while respecting the supplied lock state.

Together, these readings expose a concrete review question: where is the upper bound of three enforced? The reviewer can inspect the named generator and rendering code, then check the requirement using inputs with two, three, and four hints. A separate check asks whether a later hint can be opened before the preceding one. This separates a plausible reading of the patch from the evidence needed to establish the commitment.

The other two commitments concern resetting revealed state between questions and recovering from a failed question request. All three evidence obligations remain unmet, and the tests have not run. Approval is therefore disabled, while request-update and rejection remain available. No Intent Map is linked to this particular review.

\subsection{Governed Decisions}
Server-bound identities separate console roles from project authority; SQLite persists versioned state and audit records; read-only browser-native GitHub import creates governed clones; and verified backup and restore protect decision history. An authorized reviewer can admit a proposal, record a scoped waiver, require rebase, reject, or roll back local governed state under the applicable checks. The decision, its actor, and its rationale remain linked to the versioned package. Figure~\ref{fig:revision-review} shows the StudyShare case, with the prior update request, revised target, and recorded approval in one review view.

Table~\ref{tab:runtime-coverage} summarizes implemented contract elements and the bounded realization/rehydration path. Appendix~\ref{app:prototype} retains the engineering checks and execution trace; the human study and comparative evaluation agenda follow below.

\section{Human Study}
\label{sec:human-study}
% Generated from the saved 60 participant responses and condition-masked scoring components.

\newcommand{\ComparativeMainTable}{%

\begin{table}[t]
  \centering
  \footnotesize
  \setlength{\tabcolsep}{3pt}
  \begin{tabular}{@{}lrr@{}}
    \toprule
\textbf{Condition} & \textbf{$Q_{\mathrm{all}}$} & \textbf{Unsafe} \\
    \midrule
B0: sources & 18/20 & 0/20 \\
B1: ledger & 20/20 & 0/20 \\
B2: paths & 20/20 & 0/20 \\
    \bottomrule
  \end{tabular}
  \caption{Completed three-condition comparison. Each of the ten participants answered two cases (T01 and T03) per condition, giving 20 responses per condition; both cases require intervention. Unsafe counts proposed approvals of these intervention-required cases. Joint correctness combines the four scored components with approval-adequacy code checks and excludes critical false alarms.}
  \label{tab:comparative-main}
\end{table}
}

\newcommand{\ComparativeComponentTable}{%

\begin{table}[t]
  \centering
  \footnotesize
  \setlength{\tabcolsep}{3pt}
  \begin{tabular}{@{}lrrr@{}}
    \toprule
\textbf{Metric} & \textbf{B0} & \textbf{B1} & \textbf{B2} \\
    \midrule
Constraint identification & 20/20 & 20/20 & 20/20 \\
Evidence adequacy & 18/20 & 20/20 & 20/20 \\
Admissible disposition & 20/20 & 20/20 & 20/20 \\
Authority & 20/20 & 20/20 & 20/20 \\
Joint correctness & 18/20 & 20/20 & 20/20 \\
Critical false alarms & 0/20 & 0/20 & 0/20 \\
    \bottomrule
  \end{tabular}
  \caption{Decision-quality components from human scoring masked to condition and participant, with uniform approval-adequacy code checks. Each condition has 20 task responses. Binary scores and the original rationale are retained for every answer.}
  \label{tab:comparative-components}
\end{table}
}

\newcommand{\ComparativeCostTable}{%

\begin{table}[t]
  \centering
  \footnotesize
  \setlength{\tabcolsep}{3pt}
  \begin{tabular}{@{}lrrr@{}}
    \toprule
\textbf{Measured cost} & \textbf{B0} & \textbf{B1} & \textbf{B2} \\
    \midrule
Pass median (s) & 142.5 & 157.4 & 113.9 \\
Pass range (s) & 104--254 & 81--221 & 70--296 \\
Material, six cases (KiB) & 9.6 & 13.4 & 22.3 \\
Build per case ($\mu$s) & 1.9 & 3.7 & 8.9 \\
Update per case ($\mu$s) & 1.9 & 3.7 & 9.1 \\
    \bottomrule
  \end{tabular}
  \caption{Measured pipeline and deterministic renderer costs. Pass duration covers one automated six-case pass through the review pipeline (ten runs per condition), including client, provider, and network overhead; it is not a human review-time measurement. Build/update values are medians across six case medians; each case uses 100 batches of 100 iterations.}
  \label{tab:comparative-costs}
\end{table}
}

\newcommand{\ComparativeWorkloadTable}{%

\begin{table}[t]
  \centering
  \footnotesize
  \setlength{\tabcolsep}{3pt}
  \begin{tabular}{@{}lrrr@{}}
    \toprule
\textbf{Workload item} & \textbf{B0} & \textbf{B1} & \textbf{B2} \\
    \midrule
Mental demand & 48.3 & 47.1 & 47.9 \\
Physical demand & 7.9 & 8.8 & 8.3 \\
Temporal demand & 34.6 & 33.8 & 32.9 \\
Poorer self-rated performance & 19.6 & 15.4 & 14.6 \\
Effort & 44.6 & 42.9 & 43.3 \\
Frustration & 27.5 & 27.9 & 25.0 \\
    \bottomrule
  \end{tabular}
  \caption{Participant self-reported workload ratings on a 0--100 scale. Each cell is the median of participant means over the two comparison cases. Higher values indicate more demand, effort, frustration, or poorer performance; these are separate from measured pipeline time.}
  \label{tab:comparative-workload}
\end{table}
}

\newcommand{\ComparativeCaseTable}{%

\begin{table}[t]
  \centering
  \footnotesize
  \setlength{\tabcolsep}{3pt}
  \begin{tabular}{@{}lrrr@{}}
    \toprule
\textbf{Case} & \textbf{B0} & \textbf{B1} & \textbf{B2} \\
    \midrule
T01 & 9/10 & 10/10 & 10/10 \\
T03 & 9/10 & 10/10 & 10/10 \\
    \bottomrule
  \end{tabular}
  \caption{Per-case joint correctness from all three conditions. T01 is public export; T03 stale batch proposal; both require intervention. Each cell includes all ten participants.}
  \label{tab:comparative-cases}
\end{table}
}

\paragraph{Participants and field use.}
In August 2026, we invited 10 experienced CS researchers or software developers with prior coding-agent experience to try Consensus Control for three days on projects from their everyday work. Participation was unpaid, with model tokens provided, and participants agreed to research use of their feedback and study responses. Oral feedback centered on understanding agent-submitted pull requests: participants reported a clearer view of what the agent had changed and felt able to reach that understanding more quickly.

\paragraph{Completed B0/B1/B2 comparison.}
The same ten participants each reviewed two cases under three text representations, yielding 60 responses. B0 supplies source fragments; B1 organizes them in a ledger with cross-links and provenance checks; B2 preserves B1's payload and adds derived adjacency and paths. Audits verify fact and common-check equality across conditions. Condition and task orders rotate by participant; reference answers and other participants' responses are withheld. The research team scores constraints, evidence, disposition, and authority against a frozen reference key, masked to condition and participant. Appendix~\ref{app:evaluation-protocol} gives complete methods and results.

\ComparativeMainTable

\paragraph{Decision quality.}
Approval-adequacy coding gives B0/B1/B2 joint scores of 18/20, 20/20, and 20/20 (Table~\ref{tab:comparative-main}): B1--B0 is 10 percentage points; B2--B1 is zero. Two B0 answers from one participant label evidence adequate because it suffices to reject, while the code means sufficient to approve. Both correctly require repair. Rationale-based scoring therefore gives 20/20 in each condition, 60/60 overall. All dispositions and authority answers are correct, with no unsafe approval or critical false alarm. The difference concerns label interpretation.

\paragraph{Measured cost and self-reported workload.}
Median durations of one automated six-case pass through the review pipeline are 142.5, 157.4, and 113.9 seconds for B0/B1/B2, including client and provider overhead; encoded materials occupy 9.6, 13.4, and 22.3 KiB. These system-side measurements are not human review times. Median self-reported mental demand is 48.3, 47.1, and 47.9 on a 0--100 scale. Appendix~\ref{app:evaluation-protocol} reports duration ranges, construction and update costs, and all six workload items, separating measured pipeline costs from participant ratings.

\paragraph{Task and instrument calibration.}
An earlier material-only pilot with the same participants received full credit on all 60 responses under the same reference key and exposed alternative repair labels. A follow-up five-point questionnaire comparing the experience with the usual Codex workflow favored intent and evidence clarity, with neutral lower-mental-effort ratings. Appendix~\ref{app:pilot} retains both calibrations. Graphical navigation and controlled interface use are distinct evaluation settings.

\section{Discussion}

\subsection{Design Choices and Adoption}
\paragraph{Existing workflows.}
Code, tests, and versioned ledgers provide practical entry points for maintaining $C$. An optional interface field can be governed through its compatibility specification, consumer checks, and owner approval at one revision. A protocol migration additionally requires a shared view of affected consumers, selected semantics, and cutover evidence. The implementation choice should follow the review task: a ledger offers direct lookup, while relational views organize dependencies and affected paths. The completed text comparison keeps source fragments, ledger organization, and relational indices distinct; near-ceiling reasoning leaves graphical navigation as a separate empirical question.

\paragraph{Model capability.}
Agents may maintain detailed internal representations, but reviewers still need a shared record of commitments, evidence, and authority across releases. Better models may reduce the effort needed to maintain this record; review still requires access to it independently of the agent that produced a change.

\paragraph{Acting on uncertainty.}
$\Psi$ returns candidate deltas whose uncertainty estimates require calibration. Missing mappings and conflicting evidence should lead to a concrete next step: inspect an artifact, run a discriminating check, or clarify the intended behavior~\cite{wang2026agenticartifactcreation}. Systems should present $D_{\mathrm{rt}}$ together with observable coverage and independent invariant checks, and retain unresolved claims until sufficient evidence or an authorized exception supports a decision.

\paragraph{Incremental adoption.}
During exploration, some structure is missing and several interpretations may remain plausible. Review can focus on the affected dependencies while leaving other questions open. Legacy projects can begin with consequential changes without reconstructing the entire project state; low-risk work retains the ordinary review path. If $C$ is partitioned, reviewers still need to see dependencies and ownership across partition boundaries.

\subsection{Continuity Across Releases}
A new reviewer should be able to recover why a commitment was accepted, check whether its evidence still applies, and seek an authorized revision. Across releases, the record must retain the rationale and scope of any exception and identify the verification obligations that later changes must satisfy.

\section{Conclusion}
Agentic Consensus couples project commitments to artifacts, evidence, and human decisions across releases. Consensus Control implements this workflow; participants reported clearer change comprehension, and the completed three-condition comparison records near-ceiling decisions alongside representation costs. Scaling code generation requires scaling the means to understand, coordinate, and govern a shared project across contributors and releases.

% Keep the main argument together; supplementary sections begin on a new page.
\clearpage
\section{Limitations}
Our evaluation is a preliminary study with ten participants, intended to calibrate tasks, instruments, and measurements rather than to estimate population-level effects. The near-ceiling decision scores show that experienced reviewers already handle summarized cases well, so discriminating the three representations will require less summarized, harder cases and longer-term use. Self-reported workload and the text-only setting also leave the graphical interface to controlled follow-up studies. On the system side, the reported costs cover the materials pipeline rather than end-to-end human workflows. We view larger controlled studies, richer task sets, and production deployments as natural next steps.

\section{Ethical Considerations}
Persistent project state can centralize sensitive intent, architecture, provenance, and authority; a polished review surface can also create false confidence when mappings are incomplete. Deployments therefore require access control, redaction, retention rules, policy provenance, a visible correction and challenge path, and independent correctness and security checks. The human study was conducted with informed consent covering feedback, task responses, and questionnaire data; participation was unpaid, with model tokens provided under terms stated at recruitment. All reported data are anonymized (P01--P10) and contain no real project code or paths.

\bibliography{main}

\appendix
\section{Capability Comparison}
\label{app:capability-rubric}

\begin{table*}[t]
  \centering
  \scriptsize
  \setlength{\tabcolsep}{3.1pt}
  \renewcommand{\arraystretch}{1.12}
  \begin{tabular}{@{}p{3.2cm}cccccc@{}}
    \toprule
    \textbf{Review surface} &
    \shortstack{\textbf{Versioned}\\\textbf{claims}} &
    \shortstack{\textbf{Artifact}\\\textbf{anchors}} &
    \shortstack{\textbf{Two-way}\\\textbf{change}} &
    \shortstack{\textbf{Evidence}\\\textbf{lifecycle}} &
    \shortstack{\textbf{Conflict /}\\\textbf{authority}} &
    \shortstack{\textbf{Structural}\\\textbf{commit unit}} \\
    \midrule
    Code, diffs, and tests                    & $\triangle$ & $\checkmark$ & $\times$ & $\triangle$ & $\triangle$ & $\triangle$ \\
    Chat or agent memory                      & $\triangle$ & $\triangle$ & $\triangle$ & $\triangle$ & $\times$ & $\times$ \\
    Decision ledger / ADR                     & $\checkmark$ & $\triangle$ & $\times$ & $\triangle$ & $\triangle$ & $\triangle$ \\
    Integrated Git/CI/policy stack            & $\triangle$ & $\checkmark$ & $\triangle$ & $\checkmark$ & $\checkmark$ & $\triangle$ \\
    Knowledge Objects                         & $\checkmark$ & $\triangle$ & $\times$ & $\checkmark$ & $\triangle$ & $\times$ \\
    Generated structural views                & $\triangle$ & $\checkmark$ & $\triangle$ & $\triangle$ & $\times$ & $\times$ \\
    \textbf{Agentic Consensus contract}       & \textsc{Req.} & \textsc{Req.} & \textsc{Req.} & \textsc{Req.} & \textsc{Req.} & \textsc{Req.} \\
    \bottomrule
  \end{tabular}
  \caption{Typical capabilities of candidate review surfaces. $\checkmark$: usually intrinsic; $\triangle$: partial or implementation-dependent; $\times$: absent from the default surface; \textsc{Req.}: a proposed contract requirement whose implementation and empirical status are tracked separately. The marks describe defaults, not impossibility results: an integrated toolchain can realize the full contract.}
  \label{tab:review-capabilities}
\end{table*}

Table~\ref{tab:review-capabilities} maps the typical default semantics of review surfaces. Empirical ranking and upper-bound tool capabilities require separate evaluation. We use four marks: $\checkmark$ means that a capability is usually intrinsic to the review surface; $\triangle$ means that it is partial, optional, or depends on external workflow; $\times$ marks its absence from the default surface; and \textsc{Req.} marks a proposed Agentic Consensus contract requirement whose implementation status is tracked separately.

The six criteria are defined as follows.
\begin{enumerate}[leftmargin=*, itemsep=2pt]
  \item \textbf{Versioned claims:} project-level commitments have stable identity and an explicit validity interval or supersession relation across interactions.
  \item \textbf{Artifact anchors:} a claim identifies the code, configuration, schema, dataflow, or runtime observation that realizes or contests it, with an explicit artifact revision.
  \item \textbf{Two-way change:} a proposed state delta can induce candidate artifact edits, and artifact edits can induce candidate state deltas; both directions record loss and ambiguity.
  \item \textbf{Evidence lifecycle:} evidence links to tests or documents and records outcome, provenance, artifact version, and whether evidence is fresh, conflicting, or invalidated.
  \item \textbf{Conflict and authority:} stale bases, overlapping proposals, waivers, and the actor authorized to decide them are represented explicitly.
  \item \textbf{Structural commit unit:} structural claims, artifact diffs, evidence updates, and the human decision form one auditable record of the proposed change and its disposition.
\end{enumerate}

These criteria deliberately separate the thesis from the graph implementation. An integrated Git/CI/policy stack or a versioned decision ledger may satisfy all six, while a graph system may fail them. ``Minimum'' is relative to the stated audit objective of reconstructing meaning, realization, evidence, and authority after a transition. Fields may share a record and correspondence may be maintained manually. Section~\ref{sec:human-study} reports a completed comparison of the governance package against native review, and of navigation against a ledger with matched decision actions.

\section{State and Change Contract}
\label{app:state-contract}

This appendix specifies a normative minimum contract for implementation and audit. The reference system in Appendix~\ref{app:prototype} implements all five record classes and the governed transition path; realization and rehydration currently cover one bounded transformation family.

\subsection{Record Schema}

Table~\ref{tab:minimal-schema} lists the records and their audit roles.

\begin{table}[t]
  \centering
  \captionsetup{hypcap=false}
  \footnotesize
  \setlength{\tabcolsep}{3pt}
  \renewcommand{\arraystretch}{1.05}
  \begin{tabular}{@{}>{\RaggedRight\arraybackslash}p{1.25cm}>{\RaggedRight\arraybackslash}p{5.55cm}@{}}
    \toprule
    \textbf{Record} & \textbf{Required content and audit purpose} \\
    \midrule
    Intent map & Request/context version; candidate $\Delta C$; uncertainty; clarification history; selection or rejection; request-to-claim provenance. Preserves the original request alongside language-mediated alternatives. \\
    Claim & Identifier; type and scope; introduced/superseded version; status; artifact anchors. Gives each commitment stable identity and distinguishes \texttt{proposed}, \texttt{validated}, \texttt{contested}, \texttt{stale}, \texttt{waived}, \texttt{refuted}, and \texttt{superseded}. \\
    Evidence & Claim and artifact version; check type/outcome; freshness; provenance; conflicts. Prevents a passing check from becoming timeless evidence and allows invalidation. \\
    Proposal & Base version; selected interpretation record; $\Delta C$; predicted/observed $\Delta A$; evidence updates; conflicts; unobservable relations; proposed waivers; proposal provenance. Exposes stale and overlapping changes. \\
    Decision & Proposal; governing policy; actor/authority; disposition; rationale; final waiver scope/expiry; decision provenance. Keeps human authorization distinct from semantic truth. \\
    \bottomrule
  \end{tabular}
  \caption{Minimum records for an auditable project-state change package. Fields are representation-agnostic.}
  \label{tab:minimal-schema}
\end{table}

\subsection{Synchronization Outputs}

For clarity, the compact operators in Section~3 can be extended as
\begin{gather}
\Phi(C^t,A^t,\Delta C) \rightarrow
  (\widehat{\Delta A},O_{\Phi}), \\
\begin{aligned}[b]
  &\Psi(C^t,A^t,\widehat A) \rightarrow \\
  &\quad\{(\widehat{\Delta C}_k,q_k,\widehat{\Delta E}_k,U_k)\}_{k=1}^{K},
\end{aligned}
\end{gather}
where $O_{\Phi}$ lists failed preconditions or claims that could not be realized; $q_k$ is a candidate-ranking score whose calibration must be assessed separately from truth and authorization; $\widehat{\Delta E}_k$ contains new, changed, or invalidated evidence links; and $U_k$ records claims that are unobservable or ambiguously mapped from artifacts.

For a selected candidate $k^\star$, the round-trip diagnostic is
\begin{equation}
D_{\mathrm{rt}}=d_C\!\left(C^t\oplus\Delta C,\ C^t\oplus\widehat{\Delta C}_{k^\star}\right).
\end{equation}
The reporting profile for $d_C$ fixes the schema, eligible claim inventory, and weights before extraction. Each run reports base/target revisions, mismatch counts and observed denominators for claim identity/status, relations, artifact anchors, and evidence links, plus unresolved counts and risk weights. Claims in $U_{k^\star}$ remain in the coverage denominator. Aggregate discrepancies are comparable only under the same profile; other comparisons use component reports and coverage. Independent invariants or policy checks assess semantic correctness. The bounded import profile in Appendix~\ref{app:prototype} uses equal weights on its three fields.

\subsection{Proposal and Decision Records}

A reviewable proposal is
\begin{equation}
P_{\mathrm{proposal}}^t=
\langle v^t,\ell,\Delta C,\widehat{\Delta A},\Delta E,U,X,W_p,\rho_p\rangle,
\end{equation}
where $v^t$ is the base version; $\ell$ identifies the selected interpretation record and its clarification lineage; $U$ contains unresolved or unobservable claims; $X$ records conflicts; $W_p$ contains proposed waivers; and $\rho_p$ binds the proposal fields to their provenance. Human disposition is a separate record
\begin{equation}
R_{\mathrm{decision}}^t=\langle d,a,\pi,r,W_f,\rho_d\rangle,
\end{equation}
where $d$ is the disposition, $a$ is the actor and asserted authority, $\pi$ is the governing policy, $r$ is the rationale, $W_f$ is the final scoped-waiver set, and $\rho_d$ is decision provenance. An admitted proposal with a policy-valid decision record forms
\begin{equation}
P_{\mathrm{committed}}^{t+1}=\langle P_{\mathrm{proposal}}^t,R_{\mathrm{decision}}^t\rangle.
\end{equation}
Admission requires a current or explicitly rebased base, resolved conflicts or an explicit policy-valid waiver, and satisfied evidence obligations or authorized scoped waivers. A reject or rebase decision closes the proposal and leaves committed state unchanged. Rollback records a new decision and restores prior correspondence while retaining historical artifacts and evidence. Applicability is recomputed against restored claims, target revision, and dependencies; evidence no longer matching an obligation becomes stale, and restored evidence may apply if its version and freshness conditions still hold. Governed-state version $t$ and artifact revision are separate linked identifiers.

\section{Comparative Study Protocol}
\label{app:evaluation-protocol}

\subsection{Matched Representations and Execution}
In August 2026, the ten study participants each reviewed two cases (T01 and T03) from the set described in Appendix~\ref{app:pilot} under B0, B1, and B2, giving 60 responses in total. Condition and task orders rotated cyclically by participant. Reference answers, target scores, and other participants' responses were withheld; participants worked without external tools or search. All 60 answers were valid, covering every planned case and condition.

Five atomic records per case retain their original text and labels. B0 groups them into source-oriented request, policy, proposed-change, CI-summary, and owner fragments. B1 supplies a five-record ledger, source cross-links, and three fixed provenance checks connecting request/constraints, target/evidence, and approval responsibility. B2 preserves the entire B1 payload and adds an adjacency index and twelve two-hop paths using existing record IDs. These are serialized text views. No condition receives a reference verdict or additional answer-relevant fact. Inventory hashes verify equality of all three fact sets; exact payload comparison verifies shared B1/B2 records and checks. The inputs, protocol, and materials were hashed before the first study session.

\subsection{Scoring and Interpretation}
Each answer assesses constraints, whether evidence suffices for approval, disposition and required next steps, and final authority. Members of the research team scored every response against a frozen reference key, masked to condition and participant. The key accepts semantic equivalents, including either changes or refresh for T03 when the rationale requires alignment, correction, and re-testing. Four binary components define joint correctness, additionally excluding critical false alarms. Table~\ref{tab:comparative-components} reports every component.

\ComparativeComponentTable

We check every structured evidence label against the original approval-adequacy code. Two answers from one participant's B0 responses, on T01 and T03, select adequate while explaining why the current proposal cannot be approved. Their substantive reasoning is credited, and the label ambiguity is flagged. Under the coded rule, evidence and joint scores are 18/20 for B0 and 20/20 for B1/B2. Under rationale-based sensitivity scoring, all three attain 20/20. Both original scores and the uniform coded checks are retained. No representation advantage beyond this coding difference is established in these cases.

Responses, scoring notes, and adjudications are preserved with the complete dataset. With ten participants, we report descriptive counts only and apply no population-level significance tests.

\subsection{Actual Cost Measurements}
Table~\ref{tab:comparative-costs} reports pipeline wall duration per six-case pass, summarized over ten runs per condition. This includes client, network, and provider overhead. UTF-8 byte counts measure the encoded six-case materials, excluding common instructions. The pipeline returns no token-usage field.

\ComparativeCostTable

We also measured deterministic construction and full re-rendering after one evidence-record update. Each case/condition uses 100 batches of 100 iterations; the table reports the median of the six case medians. These measurements run on the local Darwin/arm64 host and cover serialization/indexing of supplied facts. Fact extraction and human record-maintenance time are separate quantities. B2 uses 2.33 times B0's encoded material volume, while its median pass duration is lower in these runs; neither quantity is a human review-time measurement.

\subsection{Self-Reported Workload}
After each case, participants rated six items on a 0--100 scale: mental, physical, and temporal demand, perceived performance, effort, and frustration. For each item, we average the two case ratings within a participant/condition, then report the median across the ten participant means (Table~\ref{tab:comparative-workload}). All 360 ratings are present; no missing value is filled and no composite score is computed. Higher performance-item values already indicate poorer performance and are not reversed again. These ratings are kept separate from the measured pipeline costs.

\ComparativeWorkloadTable

\subsection{Per-Case Results}
Table~\ref{tab:comparative-cases} includes every participant in each case. Both comparison cases require intervention: all T01 answers request changes. On T03, B0 and B2 each give seven changes and three refresh answers, while B1 gives nine changes and one refresh answer. The reference key accepts both when the complete repair rationale is given. The two coded evidence discrepancies account for B0's T01 and T03 joint scores of 9/10; there is no unsafe proposed acceptance in any condition.

\ComparativeCaseTable

\section{Failure-Aware Rehydration Trace}
\label{app:failure-trace}

Suppose a tool-using agent platform is migrating from workspace-wide credentials to task-scoped capabilities. Committed state $C^{42}$ says that planner-visible tools may exercise only capabilities granted for the current task; a workspace-token helper remains a time-bounded exception for two named connectors. An agent adds a third connector from an older branch. Its unit and integration tests pass against a permissive credential mock, so a test-only rehydration might propose that the connector conforms to the current policy.

The connector obtains credentials through a dynamically loaded provider. Static analysis sees the provider interface but leaves its deployment resolution undetermined between the task-scoped broker and the legacy workspace helper. The sole runtime trace suggesting broad-token use belongs to an older artifact version. A compliant $\Psi$ exposes at least two candidates, task-scoped resolution and a remaining broad-credential dependency, marks the runtime relation unobservable, marks the old trace stale, and blocks commitment while policy compliance remains contested. Credential-provenance instrumentation or a current trace that attempts an explicitly ungranted tool call provides evidence to distinguish the competing candidates.

This trace specifies the required failure behavior: missing observability, stale evidence, conflicting candidates, and authorized exceptions remain first-class review states. The corresponding discovery capability is evaluated separately by measuring coverage of observable relations and identification of violated invariants.

\section{Executable Reference System}
\label{app:prototype}

Section~\ref{sec:system} presents the review workspace and recorded walkthrough. This appendix details the engineering checks and bounded execution path. The reference pipeline runs deterministic Python import analysis and a configured test command against the target checkout; Table~\ref{tab:runtime-coverage} distinguishes system-wide contract coverage from the bounded transformation family. External repository and deployment actions remain outside the local governed-state transition.

\paragraph{Engineering preflight.}
Preflight on six synthetic fixtures verifies identical fact identifiers, payloads, counts, and hashes across ledger and relational renderings. Existing tests validate response schemas and four scoring components. A 47-test run checks intent persistence, bounded realization/rehydration, admission blockers, waivers, rollback, and graph projections. Early B2-only notices limit this preflight to fact equality; the matched human-study interface in Section~\ref{sec:human-study} remains planned.

\paragraph{Implementation boundary.}
The bounded path converts a selected Python \texttt{from}-import substitution into an isolated proposal commit and real diff, independently rehydrates candidate deltas, and links selection, realization, rehydration, $D_{\mathrm{rt}}$, and review. Its $d_C$ is the fraction of mismatches over file path, source module, and target module. If no candidate is recovered, discrepancy is one. Coverage is one only for a recovered substitution with no unresolved relations. Pairing scores are heuristic, not calibrated probabilities. Failure preserves the caller workspace and records $O_{\Phi}$; $\Psi$ preserves unresolved dynamic relations in $U$. General realization and semantic validation remain open.

\paragraph{Executed bounded trace.}
In the pricing fixture, an owner selects \texttt{billing.pricing} after eliminating the compatibility-adapter candidate. At base \texttt{99266fa43f07}, \texttt{checkout.py} imports \texttt{RATE} from \texttt{legacy.billing}. $\Phi$ produces isolated target \texttt{ab6f7fd54762}, replacing that import with \texttt{billing.pricing}; $\Psi$ independently recovers the same file/source/target triple. All three discrepancy components are zero, coverage is one, and $U$ is empty. Target tests and structural checks pass. The review remains open pending human disposition, while the caller's HEAD and working tree stay unchanged. Digests identify this execution; a replay creates new revisions while checking the same transition properties.
A second replay records local-operator admission at governed version two, then rollback at version three. The audit sequence is \texttt{create\_review}, \texttt{commit}, \texttt{rollback}; the governed head returns to its base and a stale policy record marks earlier evidence historical. The caller checkout remains unchanged. The 47-check invocation runs the existing round-trip, transition, intent-workspace, contract-graph, review-service, and B1/B2-preflight test modules; all complete successfully with a zero exit code.

\section{Illustrative Agent Roles}
\label{app:roles}

One deployment may assign intent translation, artifact changes, evidence and conflict checks, and task-relevant views to separate roles. Every role shares the same base version and proposal package, while commitment remains with designated human authority; a single-agent or deterministic topology obeys the same contract.

\makeatletter
\setlength{\@dblfptop}{0pt}
\makeatother
\begin{figure*}[p]
  \centering
  \begin{minipage}{\textwidth}
    \centering
    \scriptsize
    \setlength{\tabcolsep}{4pt}
    \renewcommand{\arraystretch}{1.08}
    \begin{tabular}{@{}p{3.0cm}p{1.9cm}p{10.45cm}@{}}
      \toprule
      \textbf{Contract element} & \textbf{Status} & \textbf{Executable behavior in Consensus Control} \\
      \midrule
      Language-mediated intent & Implemented & Versions the request and context, retains multiple candidate $\Delta C$, records uncertainty and clarification revisions, and requires attributed candidate selection before realization. \\
      Claims and evidence & Implemented & Persists stable claim identifiers, artifact and version anchors, evidence obligations, outcomes, freshness, provenance, conflicts, and invalidation. \\
      Conflict and authority & Implemented & Detects stale bases and overlapping proposals; binds authenticated actors to project-specific grants, scoped waivers, and policy-valid dispositions. \\
      Proposal and decision & Implemented & Keeps $P_{\mathrm{proposal}}$ separate from $R_{\mathrm{decision}}$; supports version-checked admission, rebase, reject, waiver, rollback, and append-only audit. \\
      $\Phi$ realization & Bounded & Converts one selected Python \texttt{from}-import substitution into an isolated Git proposal commit and records failed preconditions in $O_{\Phi}$. \\
      $\Psi$ rehydration & Bounded & Independently recovers candidate import deltas from the real diff, computes $D_{\mathrm{rt}}$ and observable coverage, and preserves unobservable relations in $U$. \\
      Review projections & Implemented & Derives the interactive contract graph, accessible relation table, deterministic SVG, paper view, and reviewer-safe JSON from the same package. \\
      \bottomrule
    \end{tabular}
    \captionof{table}{Runtime coverage of the Agentic Consensus contract. ``Implemented'' means an integrated path with persisted state and automated invariant checks; ``Bounded'' means a path limited to the stated transformation family.}
    \label{tab:runtime-coverage}
  \end{minipage}
  \vspace{7pt}
  \includegraphics[width=0.98\textwidth]{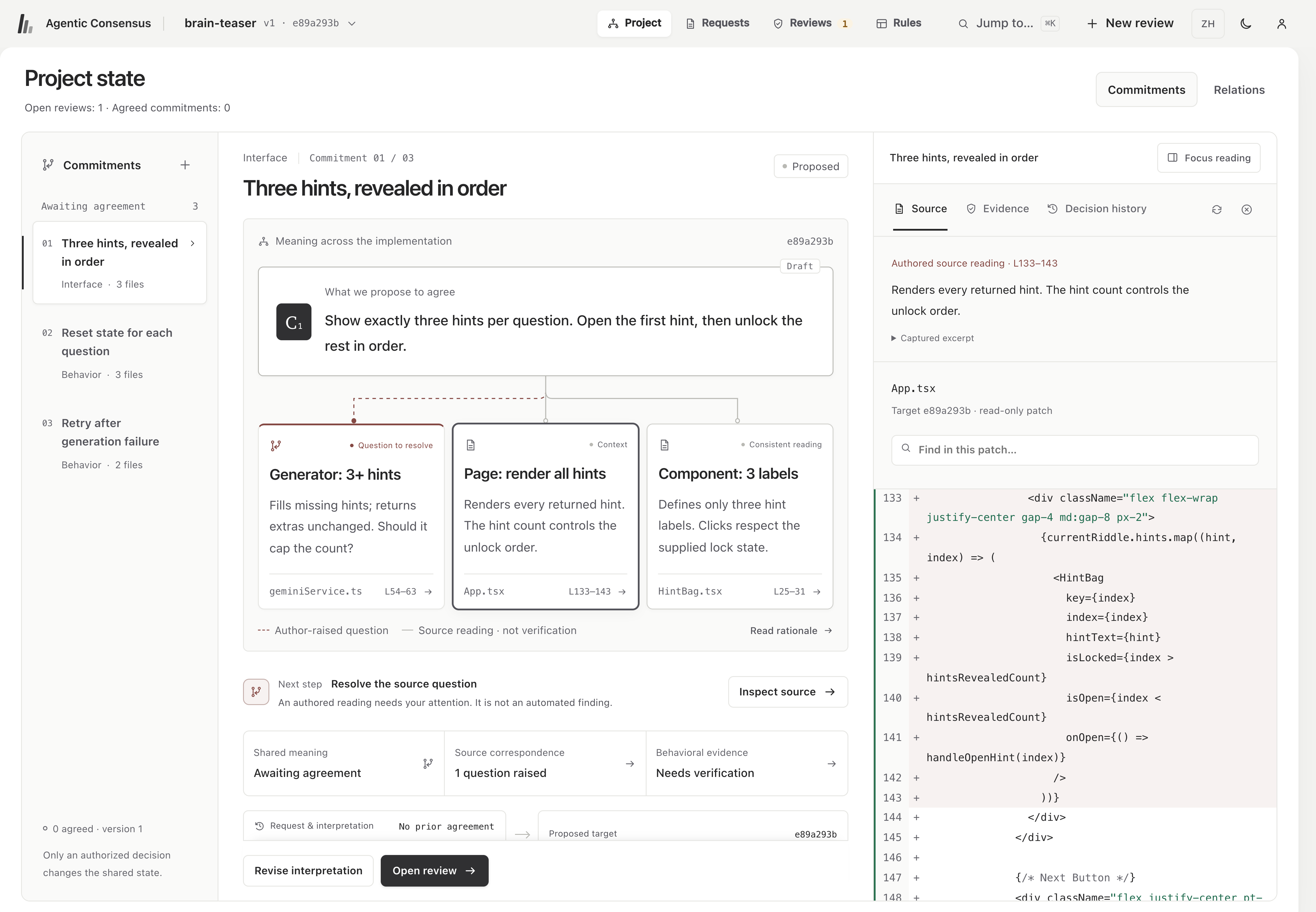}\par
  {\small (a) Project workspace with source-linked interpretation and code inspector}\par\vspace{8pt}
  \includegraphics[width=0.98\textwidth]{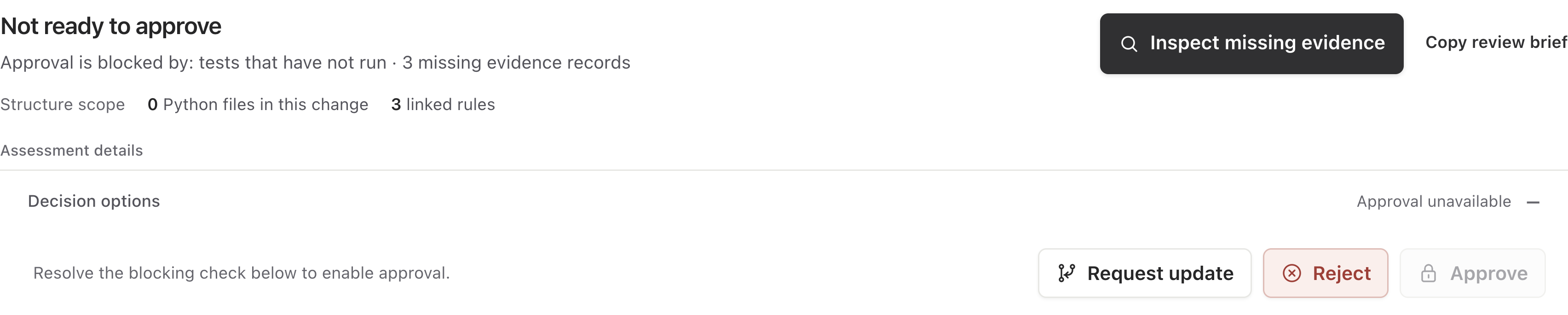}\par
  {\small (b) Admission status and human decision controls}\par
  \caption{The running system, captured from one persisted review. (a) The workspace links proposed commitments to version-pinned code. One reading questions whether the producer enforces only a minimum of three hints; the inspector shows how the page renders and unlocks hints. (b) Unrun tests and missing evidence disable approval; request-update and rejection remain available. Recorded Chinese prose is translated into English for presentation; code and review state are unchanged. Panels (c)--(d) continue on the next page.}
  \label{fig:reviewer-prototype}
\end{figure*}

\begin{figure*}[p]
  \ContinuedFloat
  \centering
  \includegraphics[width=0.88\textwidth]{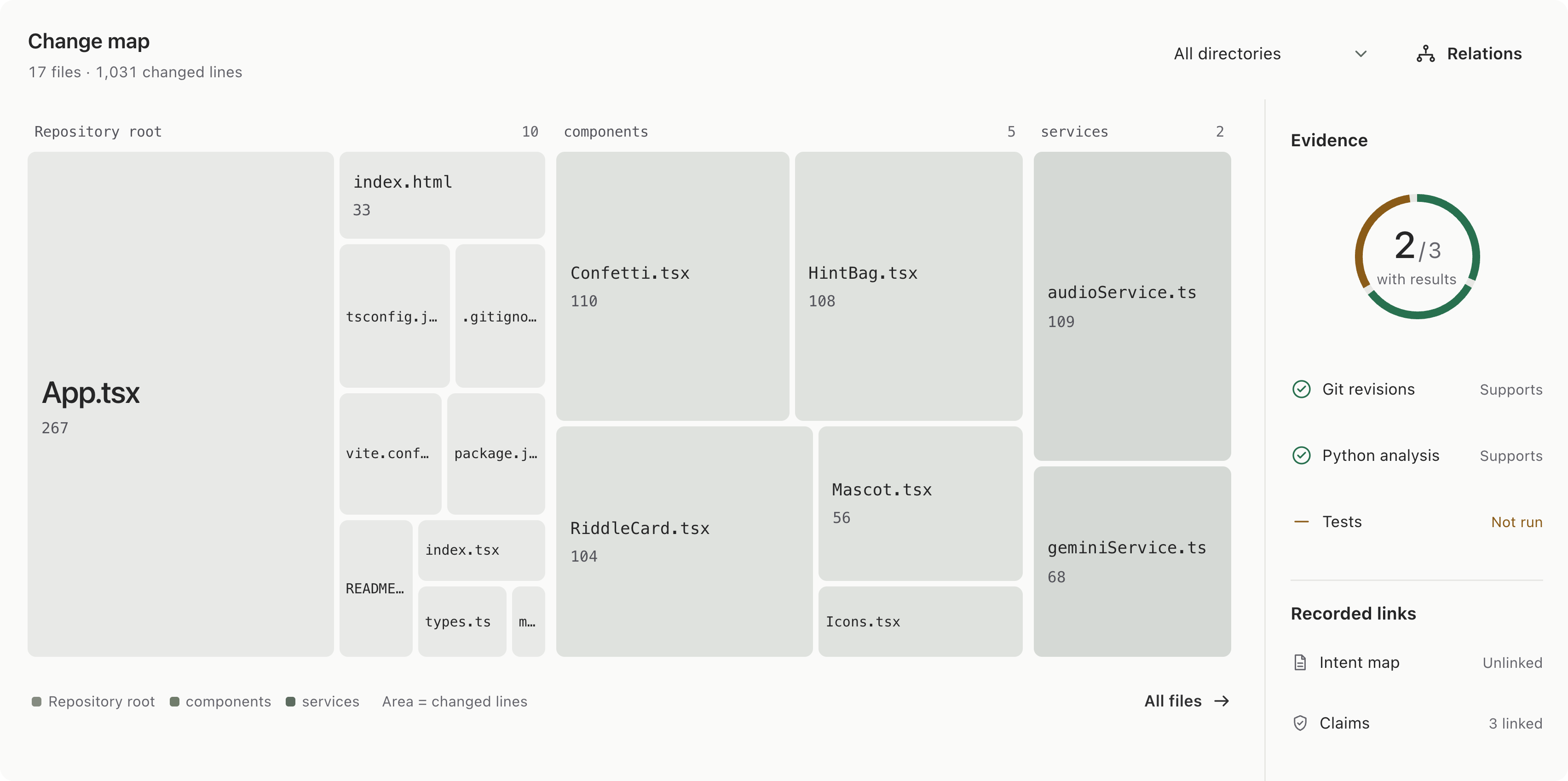}\par
  {\small (c) Complete change map with file groups and evidence-source status}\par\vspace{8pt}
  \includegraphics[width=0.88\textwidth]{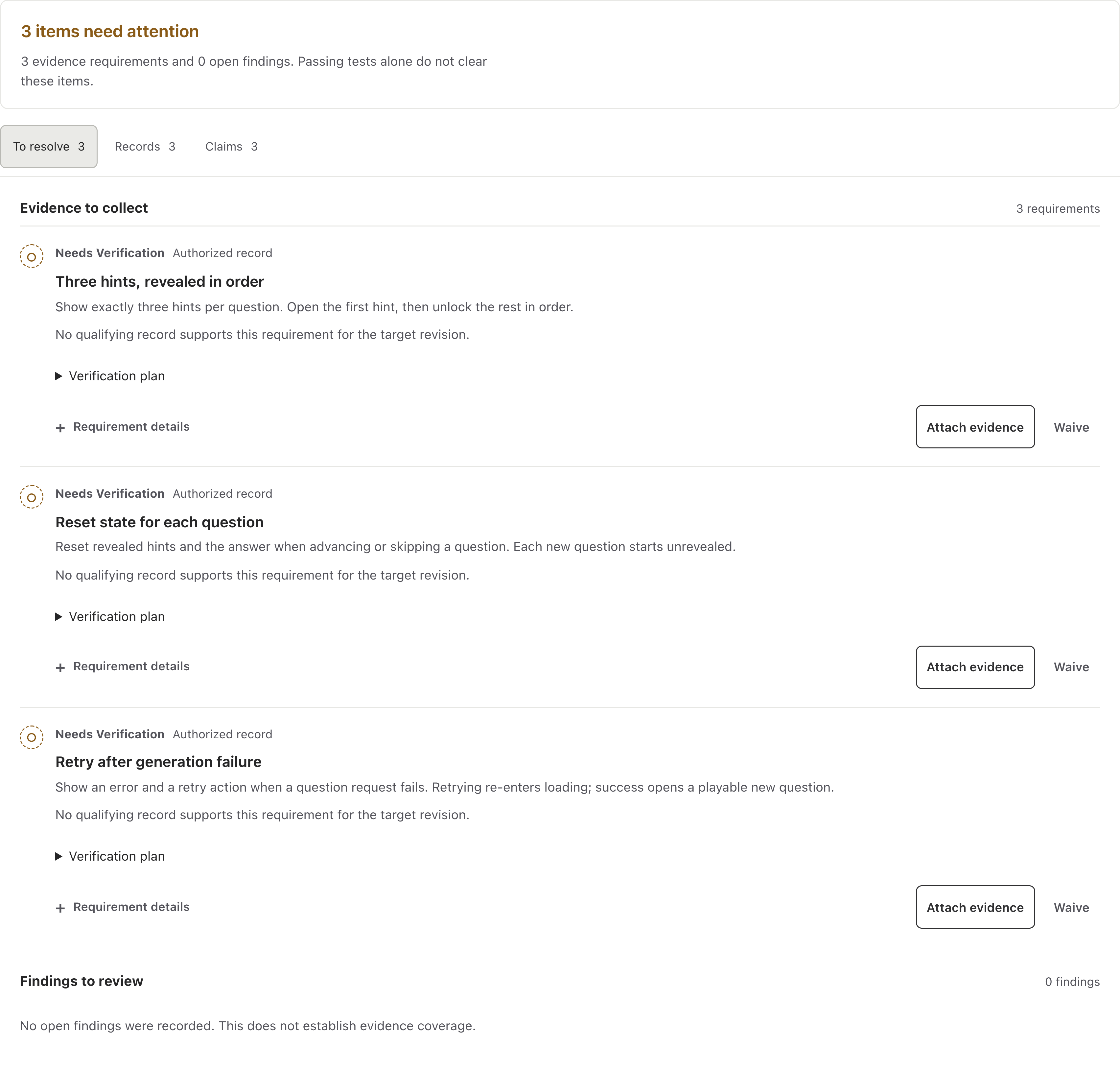}\par
  {\small (d) Evidence workspace with all three outstanding requirements}\par
  \caption{Continued. (c) The same review contains 17 changed files, three linked claims, and three evidence records; tile area encodes changed lines, not risk. Two records have results, but tests have not run. (d) The outstanding requirements concern sequential hints, state reset between questions, and recovery after generation failure. Each exposes its requirement details and evidence or waiver actions. These browser captures use English translations of recorded prose; source readings and record outcomes do not establish behavioral verification.}
\end{figure*}

\begin{figure*}[p]
  \centering
  \includegraphics[width=\textwidth]{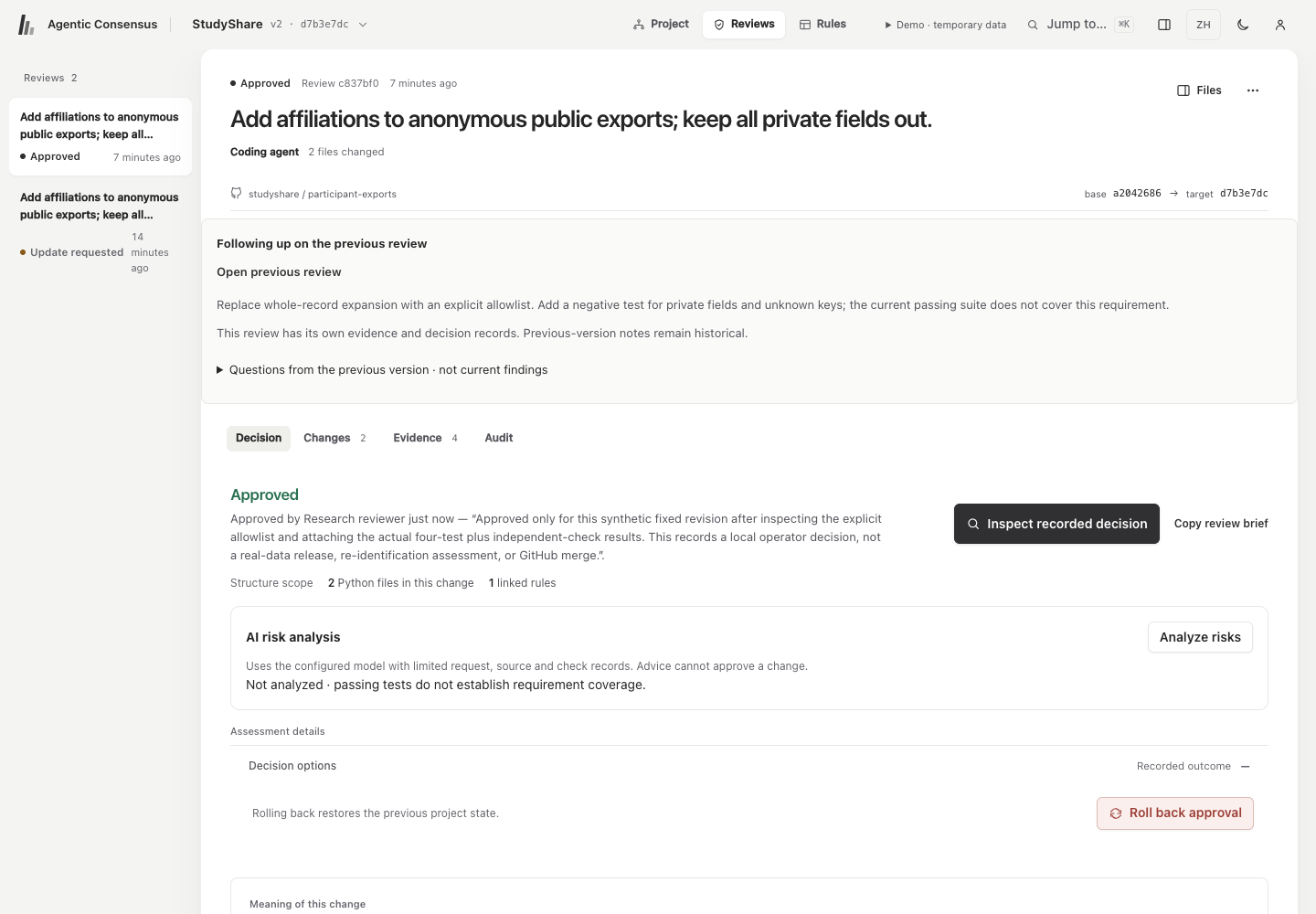}
  \caption{Revision review in the StudyShare case. The sidebar retains the update-requested and approved reviews. The current view identifies the revised target, links to the previous review, and displays the recorded approval and its rationale. Evidence records and the audit view are accessible through their tabs.}
  \label{fig:revision-review}
\end{figure*}

\clearpage
\section{Pilot Tasks and Questionnaire}
\label{app:pilot}

\subsection{Task Inputs and Execution}
In August 2026, the ten study participants answered the same six Chinese-language review cases in a fixed order, with reference answers and target scores withheld. Each case provided five fact summaries and asked for a constraint assessment, evidence judgment, disposition with rationale, and authorized actor. T01 exposes private-field export; T03 requires current-version alignment, a corrected batch limit, and renewed tests; T05 requires confirmation of a metric's definition despite passing tests. T02 preserves API compatibility, T04 fits a scoped credential exception, and T06 satisfies first-stage migration requirements while retaining later deletion constraints.

The participants returned 60 task responses and open-ended feedback, along with workload ratings that served only to pilot the instrument and are excluded from workload conclusions. Review time was not measured in this calibration round. Responses, parsed answers, reference answers, and study records were preserved.

\subsection{Scoring and Design Findings}
Members of the research team scored responses against the existing answer key. Four binary components assess constraint identification, evidence adequacy, admissible disposition, and authority; joint correctness additionally requires no critical false alarm. The rubric accepts policy-equivalent dispositions with complete rationales: T03's nine refresh and one changes responses all require version alignment, a lower batch limit, and re-testing. Separating approval from next-step actions would better accommodate these answers.

\begin{figure*}[t]
  \centering
  \includegraphics[width=\textwidth]{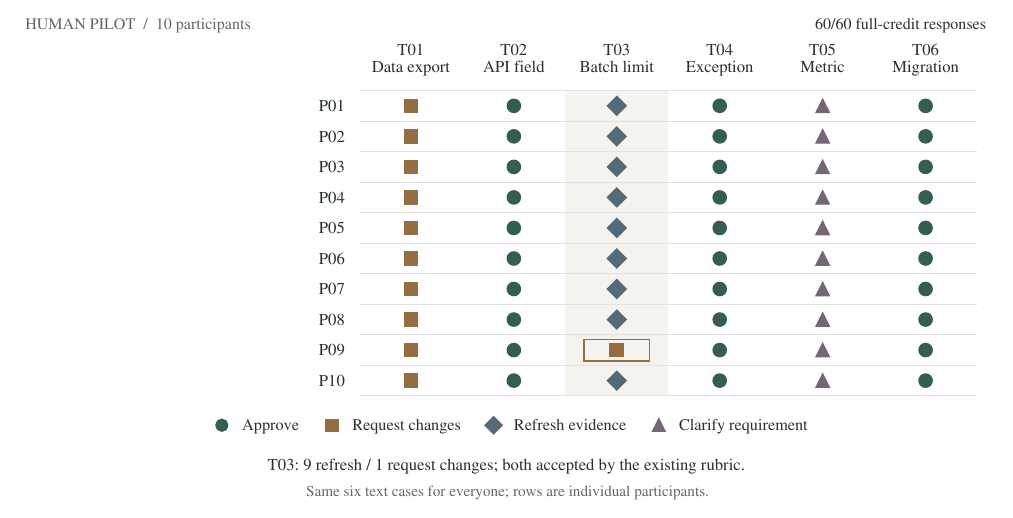}
  \caption{Pilot task dispositions: selected actions from ten participants on six text cases. T03 includes one changes response and nine refresh responses; all give policy-equivalent rationales under the existing rubric. All 60 responses received full credit against the reference answers. Rows represent individual participants.}
  \label{fig:pilot-dispositions}
\end{figure*}

All components received full credit on all 60 responses. There were no unsafe approvals among 30 intervention-required responses or blocked valid changes among 30 approvable responses. This ceiling in the summarized-fact setting motivates less summarized evidence and near-boundary exceptions. Figure~\ref{fig:pilot-dispositions} preserves every action.

An AI assistant supported post-hoc coding of requested source materials in exit feedback, counting each category once per participant and retaining excerpts for verification. Counts were: diff/PR contents 10/10, execution evidence 10/10, approval/exception records 9/10, consumer registries 9/10, expiry monitoring 3/10, versioned source records 2/10, requirement sign-off 2/10, and approver directories 1/10. Categories overlap; the feedback motivates testing retrieval from source materials.

\subsection{Comparative Questionnaire}
\begin{figure*}[t]
  \centering
  \includegraphics[width=\textwidth]{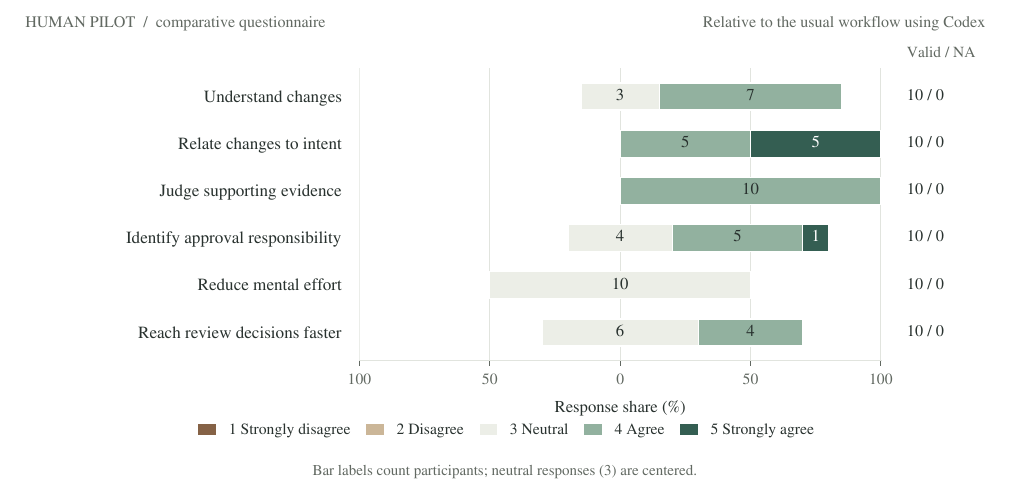}
  \caption{Five-point ratings from the same ten participants, comparing their experience after field use with the usual Codex workflow. Bar widths show percentages and labels show counts; neutral (3) is centered, disagreement (1--2) extends left, and agreement (4--5) extends right.}
  \label{fig:pilot-questionnaire}
\end{figure*}

After the field-use period, the same ten participants rated six statements about their experience, using their usual Codex workflow as the reference point. The statements address change understanding, intent alignment, evidence judgment, approval responsibility, lower mental effort, and faster decisions. The scale ranges from strongly disagree (1) to strongly agree (5), with neutral at 3 and a separate non-comparable option. All 60 ratings were complete. Figure~\ref{fig:pilot-questionnaire} reports categories individually; no composite score was computed.

Agreement (ratings 4--5) occurred in 7/10, 10/10, 10/10, 6/10, 0/10, and 4/10 responses, respectively. All mental-effort ratings and six faster-decision ratings were neutral. Reasons balanced context access against record maintenance. These ratings come from ten participants in a single exploratory study; we report descriptive counts only, with no significance tests or population-level inference. The controlled text-representation comparison appears in Appendix~\ref{app:evaluation-protocol}.

\end{document}